\documentclass{llncs}
\usepackage{wrapfig}
\usepackage{etex}
\usepackage[section]{placeins}
\usepackage[english]{babel} 
\usepackage[latin1]{inputenc} 
\usepackage{latexsym}
\usepackage{epsfig} 
\usepackage{listings}
\usepackage{url,xspace} 
\usepackage[draft]{fixme}
\usepackage{color} 
\usepackage{multirow}
\usepackage{latexsym,amssymb}
\usepackage{fancybox,epic}
\usepackage[noend]{algpseudocode}
\usepackage{amsmath}
\usepackage{graphics}
\usepackage{graphicx}
\usepackage{todonotes}
\usepackage{hyperref}
\usepackage{paralist}

\usepackage{graphicx}
\usepackage{xcolor}
\definecolor[named]{ACMBlue}{cmyk}{1,0.1,0,0.1}
\definecolor[named]{ACMYellow}{cmyk}{0,0.16,1,0}
\definecolor[named]{ACMOrange}{cmyk}{0,0.42,1,0.01}
\definecolor[named]{ACMRed}{cmyk}{0,0.90,0.86,0}
\definecolor[named]{ACMLightBlue}{cmyk}{0.49,0.01,0,0}
\definecolor[named]{ACMGreen}{cmyk}{0.20,0,1,0.19}
\definecolor[named]{ACMPurple}{cmyk}{0.55,1,0,0.15}
\definecolor[named]{ACMDarkBlue}{cmyk}{1,0.58,0,0.21}
\hypersetup{colorlinks,
  linkcolor=ACMDarkBlue,
  citecolor=ACMPurple,
  urlcolor=ACMDarkBlue,
  filecolor=ACMDarkBlue}

\usepackage{epsfig}

\usepackage{booktabs}
 \usepackage[all]{xy}

\usepackage{amssymb}
\usepackage{latexsym}
\usepackage{fancybox,epic}
\usepackage{algorithm}
\usepackage{amsmath}
\usepackage{colortbl}

\usepackage{pgf}
\usepackage{tikz}
\usepackage{amsmath}

\let\subparagraph\paragraph

\usepackage{listings}

\definecolor{deepblue}{rgb}{0,0,0.5}
\definecolor{deepred}{rgb}{0.6,0,0}
\definecolor{deepgreen}{rgb}{0,0.5,0}
\definecolor{halfgray}{gray}{0.55}
\definecolor{ipythonframe}{RGB}{207, 207, 207}

\definecolor{ckeyword}{HTML}{7F0055}
\definecolor{ccomment}{HTML}{3F7F5F}
\definecolor{cnumber}{HTML}{2A0099}
\definecolor{pblue}{rgb}{0.13,0.13,1}
\definecolor{pgreen}{rgb}{0,0.5,0}
\definecolor{pred}{rgb}{0.9,0,0}
\definecolor{pgrey}{rgb}{0.46,0.45,0.48}

\definecolor[named]{ACMBlue}{cmyk}{1,0.1,0,0.1}
\definecolor[named]{ACMYellow}{cmyk}{0,0.16,1,0}
\definecolor[named]{ACMOrange}{cmyk}{0,0.42,1,0.01}
\definecolor[named]{ACMRed}{cmyk}{0,0.90,0.86,0}
\definecolor[named]{ACMLightBlue}{cmyk}{0.49,0.01,0,0}
\definecolor[named]{ACMGreen}{cmyk}{0.20,0,1,0.19}
\definecolor[named]{ACMPurple}{cmyk}{0.55,1,0,0.15}
\definecolor[named]{ACMDarkBlue}{cmyk}{1,0.58,0,0.21}

\lstdefinelanguage{Solidity} {
  keywords={typeof, modifier, function, public, returns, external,
  contract, new, true, false, private, catch, function, return, null, throw, catch, switch, var, if, in, while, do, else, case, break},
  ndkeywords={bool, mapping, bytes32, string},
  identifierstyle=\color{black},
  sensitive=false,
  comment=[l]{//},
  morecomment=[s]{/*}{*/},
  commentstyle=\color{ccomment}\ttfamily,
  string=[b]",
  showstringspaces=false,
  morestring=[b]',
  showspaces=false,
  showtabs=false,
  breaklines=true,
  morekeywords={function, contract, returns, return},
  breakatwhitespace=true,
  lineskip=-0.6pt,
  basewidth={0.54em, 0.4em},
  basicstyle=\scriptsize\ttfamily,
  keywordstyle={\color{ckeyword}\scriptsize\bfseries},
  ndkeywordstyle={\color{black}\scriptsize\bfseries},
  commentstyle={\color{ccomment}\itshape},
  stringstyle={\color{pgreen}},
  numberstyle={\scriptsize\color{cnumber}\ttfamily},
  moredelim=[il][\textcolor{pgrey}]{$$},
  moredelim=[is][\textcolor{pgrey}]{\%\%}{\%\%},
}

\newcommand{\scode}[1]{\lstinline[language=Solidity,basicstyle=\small\ttfamily]{#1}}
\newcommand{\code}[1]{\scode{#1}}

\definecolor{shadecolor}{gray}{1.00}
\definecolor{ddarkgray}{gray}{0.75}
\definecolor{darkgray}{gray}{0.30}
\definecolor{light-gray}{gray}{0.87}

\newcommand{\tname}[1]{\textsf{#1}\xspace}

\newcommand{\plname}[1]{\textsf{#1}\xspace}

\newcommand{\EVM}{\code{EVM}\xspace}

\newcommand{\toolname}{\tname{\textsf{SAFEVM}}} 

\newcommand{\solidity}{\plname{Solidity}}
\newcommand{\vyper}{\plname{Vyper}}
\newcommand{\oyente}{\tname{Oyente}}

\newcommand{\ethir}{\tname{EthIR}}

\newcommand{\cpa}{\plname{CPAchecker}}
\newcommand{\verymax}{\plname{VeryMax}}
\newcommand{\seahorn}{\plname{SeaHorn}}

\newcommand{\lst}[1]{\lstinline!#1!}

\newcommand{\Get}[1]{\ensuremath{\mbox{\bf \lstinline!get!}}\xspace}

\newcommand{\returnval}[1]{\mbox{\lstinline!return #1!}\xspace}

\newcommand{\extend}[1]{S}

\makeatletter
\newsavebox\myboxA
\newsavebox\myboxB
\newlength\mylenA

\newcommand*\xoverline[2][0.75]{%
    \sbox{\myboxA}{$\m@th#2$}%
    \setbox\myboxB\null
    \ht\myboxB=\ht\myboxA%
    \dp\myboxB=\dp\myboxA%
    \wd\myboxB=#1\wd\myboxA
    \sbox\myboxB{$\m@th\overline{\copy\myboxB}$}
    \setlength\mylenA{\the\wd\myboxA}
    \addtolength\mylenA{-\the\wd\myboxB}%
    \ifdim\wd\myboxB<\wd\myboxA%
       \rlap{\hskip 0.7\mylenA\usebox\myboxB}{\usebox\myboxA}%
    \else
        \hskip -0.5\mylenA\rlap{\usebox\myboxA}{\hskip 0.7\mylenA\usebox\myboxB}%
    \fi}
\makeatother

\lstdefinestyle{numbers}
{numbers=left, numberstyle=\tiny}

\def\anno#1{{\ooalign{\hfil\raise.07ex\hbox{\small{\rm
          \textcolor{red}{{\tiny #1}}}}\hfil%
        \crcr{\scriptsize \textcolor{blue}{\mathhexbox20D}}}}}

\newcommand{\invalidLine}{\anno{!}}
\newcommand{\addedLine}{\anno{+}}

\title{\toolname: A Safety Verifier for Ethereum Smart Contracts}

 \author{Elvira Albert$^1$ \and Jes\'us Correas$^1$ \and  Pablo Gordillo$^1$
 \and \\ Guillermo Rom\'an-D\'iez$^2$ \and Albert Rubio$^1$ }

\institute{
  Complutense University of Madrid,  Spain \and
  Universidad Polit\'ecnica de Madrid, Spain
}

\begin{document}
\maketitle

\begin{abstract}
  Ethereum smart contracts are public, immutable and distributed and,
  as such, they are prone to vulnerabilities sourcing from programming
  mistakes of developers. 
  This paper presents \toolname
  , a verification tool for Ethereum
  smart contracts that makes use of state-of-the-art verification engines for
  C programs.
\toolname takes as input an Ethereum smart
  contract (provided either in \textsf{Solidity} source code, or in compiled
  \EVM bytecode), optionally with \textsf{assert} and \textsf{require}
  verification
  annotations, 
  and produces in the output a report with the verification
  results. 
  Besides general safety annotations, \toolname handles the
  verification of array accesses: it
  automatically generates \textsf{SV-COMP} verification assertions
  such that C verification engines can prove safety of array
  accesses. Our experimental evaluation has been undertaken on all
  contracts pulled from \textsf{etherscan.io} (more than 24,000) by
  using as back-end verifiers \textsf{CPAchecker}, \textsf{SeaHorn}
  and \textsf{VeryMax}.
\end{abstract}





\section{Overview of \toolname}


\begin{figure}[t]
\begin{center}
\includegraphics[scale=0.48]{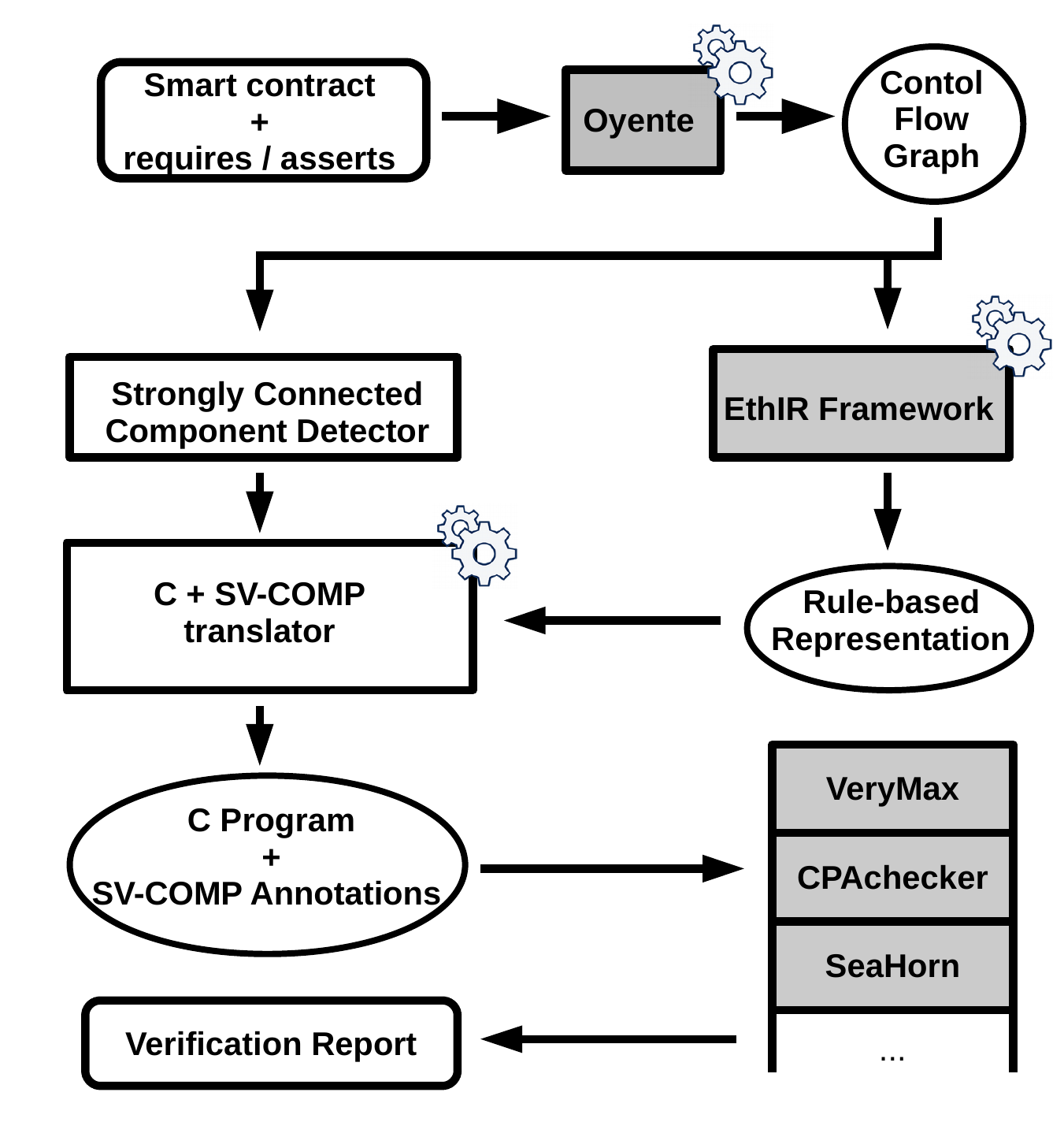} 
\caption{\toolname's architecture}  \label{tool} 
\end{center}
\end{figure}

Each blockchain provides its own programming language to implement
smart contracts. \textsf{Solidity}, a Turing complete language, is the
most popular language to write smart contracts for the Ethereum
platform that are then compiled to \EVM (Ethe\-reum Virtual Machine~\cite{yellow})
bytecode.
Each instruction executed by the \EVM has an associated gas
consumption specified by Ethereum.  Being security a main concern of
Ethereum, the \textsf{Solidity} language contains the
verification-oriented functions, \code{assert} and \code{require}, to
check for safety conditions or requirements and terminate the
execution if they are not met. 
As usual, the \code{assert}
function can be used for verification purposes (e.g., to check
invariants), while the \code{require} function is used to specify
preconditions (e.g., to ensure valid conditions on the inputs or
contract state variables, or to validate return values from calls to
external contracts). When the \solidity code is compiled into
\EVM bytecode, the \code{require} condition is transformed into a test
that checks the condition and invokes a \code{REVERT} bytecode if it
does not hold. \code{REVERT} aborts the whole execution of the smart
contract, reverts the state and all remaining gas is refunded to the caller. 
The \code{assert} checks the condition and
invokes an \code{INVALID} bytecode if it does not hold. When executing
\code{INVALID}, the state is reverted but no gas is refunded, and
hence it has more serious consequences than \code{REVERT}: besides
the economic consequences of losing the gas, the only information
given to the transaction is an out-of-gas error message.  The
treatment of array accesses is done similarly as for the
\code{assert}, when an array position is accessed, the generated \EVM
bytecode checks if the position accessed is within the array bounds
and otherwise the \code{INVALID} bytecode is executed. Division and related bytecodes like \code{MOD}, \code{SMOD},
\code{ADDMOD}, \code{MULMOD}, also lead to executing \code{INVALID} when
the denominator is zero.



Therefore, the \code{INVALID} bytecodes are key for the verification
of the Ethereum smart contracts, as they capture both assertion
violations and several sources of fatal operations (e.g.,
out-of-bounds access, division by zero).  In essence, our approach to
the verification of smart contracts consists in decompiling the \EVM
bytecode for the smart contract into a C program with \code{ERROR}
annotations (following the \code{SV-COMP}
format, \url{https://sv-comp.sosy-lab.org/2019/rules.php}) to
enable their verification using existing tools for the verification of
C programs. Developing the verifier from the low-level \EVM has
important advantages: (i) sometimes the source code is not available
(e.g., the blockchain only stores the bytecode), (ii) the
\code{INVALID} bytecodes are visible at the level of bytecode and we
can give a uniform treatment to the various safety concerns described
above, (iii) our analysis works for any other language that compiles to
\EVM (e.g., \vyper), and it is not affected by changes in the source
language, or by compiler optimizations.  Luckily, there are a number
of open-source tools that help us in the decompilation process and
that we have integrated within our tool-chain.

 Fig.~\ref{tool}
depicts the main components of \toolname that are as follows (shaded
boxes are off-the-shelf used systems not developed by us):
\begin{enumerate}
\item \emph{Input.} \toolname takes a smart contract, optionally with
  \code{assert} and \code{require} verification annotations. The
  smart contract can be given in \solidity source code or in 
 \EVM compiled code. In the latter case, the annotations have been
 compiled into bytecode as described above.
\item \emph{CFG.} In either form, the code is given to \oyente
 \cite{Oyente}, a symbolic execution engine that has been extended to
 compute the complete CFG from the given smart contract. As \oyente
 does not handle recursive functions, they are already discarded at
 this step. The CFG generation phase is not described in the paper, we
 refer to \cite{Oyente,AlbertGLRS18}.
%
\item \emph{EthIR.} The decompilation of the \EVM bytecode into a
  higher-level \emph{rule-based representation} (RBR) is carried out
  from the generated CFG by \ethir \cite{AlbertGLRS18}. Technical
  details of this phase are
  not described in the paper, we refer to \cite{AlbertGLRS18}.
\item \emph{C+SV-COMP translator.} We have implemented a translator
  for the recursive RBR representation into an \emph{abstract} Integer C
  program (i.e., all data is of type Integer) with verification
  annotations using the \code{SV-COMP} format. Features of the \EVM that we
  cannot handle yet (e.g., bit-wise operations) are abstracted away in
  the translation (see Sec.~\ref{translator}).
  \code{INVALID} instructions are transformed into
    \code{ERROR} annotations in the C program following the \code{SV-COMP} format.
\item \emph{Verification.} Any verification tool for Integer C
  programs that uses \code{SV-COMP} annotations can be used to verify the safety
  of our C-translated contracts. We have evaluated our approach using
  three state-of-the-art C verifiers, \textsf{CPAchecker}
  \cite{DBLP:conf/cav/BeyerK11}, \verymax \cite{Brockschmidt15}, and
  \textsf{SeaHorn} \cite{Kahsai15}, and the verification report they produce is
  processed by us to report the results in terms of functions of the smart
  contract.
\end{enumerate}

Our tool \toolname has a very large (potential) user base, as Ethereum
is currently the most advanced platform for coding and processing
smart contracts.
As we
will describe in Sec.~\ref{experiments}, using \toolname we have
automatically verified safety of around 20\% of \emph{all} functions 
(depending on the
verifier) that might execute \code{INVALID} bytecodes from the whole set of
contracts pulled from \textsf{etherscan.io} (more than 24,000
contracts), and we have found potential vulnerabilities in functions
that could not be verified.




\section{Translation to C with SV-COMP Annotations} 
\label{translator}


\begin{figure}[t]
  \begin{minipage}{0.5\textwidth}
  \begin{center}
  \begin{lstlisting}[name=code, numbersep=6pt]
contract SmartBillions {
 struct Wallet {
   ...,  uint16 lstDvdndPrd;} 
 uint public dvdndPrd;
 uint[] public dvdnds;
 mapping(address => uint) blncs;
 uint public ttlSpply;
 mapping (address => Wallet) wllts;

 function commitDividend(address wh) {
$\label{run:req1}$$\addedLine$  //require(ttlSpply > 0);
$\label{run:req2}$$\addedLine$  //require(dvdndPrd < dvdnds$.$length);
  uint lst=wllts[wh].lstDvdndPrd;
$\label{run:req3}$$\addedLine$  //require(dvdndPrd >= lst);
  ...
$\label{run:div1}$$\hspace{-0.2cm}\invalidLine$  uint shr=blncs[wh]*$\textsf{0xffffffff}$/ttlSpply;
  uint blnc = 0; 
$\label{run:for-init}$  for(;lst<dvdndPrd;lst++) {
$\label{run:array}$$\hspace{-0.2cm}\invalidLine$      blnc += shr * dvdnds[lst]; 
$\label{run:for-end}$  }
$\label{run:assert}$$\hspace{-0.2cm}\invalidLine$$\addedLine$  assert(lst == dvdndPrd); 
$\label{run:div2}\hspace{-0.2cm}$  blnc = (blnc/$\textsf{0xffffffff}$); 
   ...
 }
}
  \end{lstlisting}
  \end{center}
  \end{minipage}
  \begin{minipage}{0.48\textwidth}
  \begin{center}
  \begin{lstlisting}[numbers=none, 
    basicstyle=\footnotesize, 
    commentstyle=\footnotesize\textit]
block734(s$_5$,...,s$_0$,g$_4$,g$_1$,g$_0$,l$_3$,l$_2$) $\leftarrow$
  ...        // block734$ $instructions
  call(jump734(s$_7$,...,s$_0$,g$_4$,g$_1$,g$_0$,l$_3$,l$_2$))
jump734(s$_7$,...,s$_0$,g$_4$,g$_1$,g$_0$,l$_3$,l$_2$) $\leftarrow$
  geq(s$_7$,s$_6$),     // lst$\ensuremath{\geq}$dvdndPrd    
  call(block789(s$_5$,...,s$_0$,g$_4$,g$_0$,l$_3$,l$_2$))
jump734(s$_7$,...,s$_0$,g$_4$,g$_1$,g$_0$,l$_3$,l$_2$) $\leftarrow$
  lt(s$_7$,s$_6$),      // lst$\ensuremath{<}$dvdndPrd  
  call(block745(s$_5$,...,s$_0$,g$_4$,g$_1$,g$_0$,l$_3$,l$_2$))
block745(s$_5$,...,s$_0$,g$_4$,g$_1$,g$_0$,l$_3$,l$_2$) $\leftarrow$
  ...        // block745$ $instructions
  call(jump745(s$_9$,...,s$_0$,g$_4$,g$_1$,g$_0$,l$_3$,l$_2$))
jump745(s$_9$,...,s$_0$,g$_4$,g$_1$,g$_0$,l$_3$,l$_2$) $\leftarrow$
  lt(s$_9$,s$_8$),      // lst$\ensuremath{<}$dvdnds.length
  call(block759(s$_7$,...,s$_0$,g$_4$,g$_1$,g$_0$,l$_3$,l$_2$))
jump745(s$_9$,...,s$_0$,g$_4$,g$_1$,g$_0$,l$_3$,l$_2$) $\leftarrow$
  geq(s$_9$,s$_8$),     // lst$\ensuremath{\geq}$dvdnds.length
  call(block758(s$_7$,...,s$_0$))
block758(s$_7$,...,s$_0$) $\leftarrow$
  INVALID
block759(s$_7$,...,s$_0$,g$_4$,g$_1$,g$_0$,l$_3$,l$_2$) $\leftarrow$	
  // block759$ $instructions
  ...
  s$_6$ = s$_7$+s$_6$,             // ADD  
  s$_6$ = fresh$_0$,           // SLOAD
  s$_7$ = s$_4$,               // DUP3
  ...
  call(block734(s$_5$,...,s$_0$,g$_4$,g$_1$,g$_0$,l$_3$,l$_2$))
  \end{lstlisting}
  \end{center}
  \end{minipage} 
  \caption{\textsf{Solidity} code (left) and excerpt of RBR rules of
  \textsf{for} loop (lines \ref{run:for-init}-\ref{run:for-end})}
\label{fig:running}
\end{figure}

As motivating example, we use a \solidity contract that implements a
lottery system called \emph{SmartBillions} (available at
  \url{https://smartbillions.com/}).
We illustrate the safety verification of its internal function
\code{commitDividend} (an excerpt of its code appears to the left of
Fig.~\ref{fig:running}) that commits remaining dividends to the user
\code{wh}. We have shortened the variable names
by removing the vowels from the names.
Lines marked with \invalidLine~might lead to executing different
sources of \code{INVALID}: Line \ref{run:div1} (L\ref{run:div1} for
short) to a division by zero when \code{ttlSpply} is $0$; at
L\ref{run:array} when \code{lst} $\geq$ \code{dvdnds.length} and
thus it is accessing a position out of the bounds of the array; and at
L\ref{run:assert} when the condition within the assert does not
hold. In order to be able to verify its safety (i.e., absence of
\code{INVALID} executions), we add the lines marked with
\addedLine~that introduce error-handling functions \code{require} and
\code{assert} in the verification process.

The starting point of our translator 
is the RBR produced by \ethir \cite{AlbertGLRS18}.  The RBR is
composed of a set of rules containing decompiled versions of bytecode
instructions (e.g., \code{LOAD} and \code{STORE} are decompiled into
assignments) and whose structure of rule invocations is obtained from
the CFG produced by \oyente. The RBR might contain two kinds of rules:
sequences of instructions referred to as $blockX$, and conditional
jump rules, named $jumpX$, whose first instruction is the Boolean
condition used to select between the rules of the function definition.
Rule parameters include: the operand stack flattened in variables
named $s_i$, the state of the contract (this is the global data),
named $g_i$, and the local memory (represented by local variables),
named $l_i$.
To the right of Fig.~\ref{fig:running} we show the fragment of the RBR
produced by \ethir for the loop of
L\ref{run:for-init}-L\ref{run:for-end}.
At rule $block759$ we show the transformation of some \EVM bytecodes
(the original bytecodes appear in comments \code{//}) into
higher-level RBR instructions. The RBR is already \emph{abstract} in
the sense that when variables refer to state or memory locations that are not
known they become fresh variables (see variable \code{fresh}$_0$ in
$block759$) so that a posterior analysis will not assume any value for
them (details are in \cite{AlbertGLRS18}).
Observe that the fragment of the RBR contains an \code{INVALID}
instruction within $block758$ and such block can be executed when
$geq(s_9,s_8)$ (see rule $jump745$). By tracking variable assignments,
we can infer that $s_9$ contains the value of \code{lst} and $s_8$
the size of \code{dvdnds}, hence the comparison is checking
out-of-bounds array access.  The remaining of the section explains the
main four phases of the translation from the RBR to an abstract
Integer C program.


%


\begin{figure} 
\hspace{0.5cm}   \begin{minipage}{0.53\textwidth}
  \begin{center}
  \begin{lstlisting}[name=code]
$\label{runc:field1}$int g0 = __VERIFIER_nondet_int(); 
...
$\label{runc:field2}$int g4 = __VERIFIER_nondet_int(); 
$\label{runc:local1}$int l0 = __VERIFIER_nondet_int();   
...
$\label{runc:local2}$int l3 = __VERIFIER_nondet_int();  
$\label{runc:paramwho}$int who = __VERIFIER_nondet_int(); 
$\label{runc:stack1}$int s0;   
$\ldots$
$\label{runc:stack9}$int s9;  

void block758() {
$\label{runc:error}$  ERROR: __VERIFIER_error();
}
\end{lstlisting}
\end{center}
\end{minipage}
\begin{minipage}{0.45\textwidth}
\begin{center}
\begin{lstlisting}[name=code]
void block734(){
 init_loop_0:
   // block734 instructions
   if(s7 >= s6){ // jump734
     block789();
     goto end_loop_0; }
   // block745 instructions
   if(s9 >= s8){ // jump745
     block758();
     goto end_loop_0; }
  // block759 instructions
  s6 = s7 + s6
$\label{runc:sload}$  s6 = __VERIFIER_nondet_int();  
  s7 = s4;
  ...
  goto init_loop_0;
 end_loop_0:   ;}
  \end{lstlisting}
  \end{center}
  \end{minipage}
\caption{\textsf{C} translated code with \textsf{SV-COMP} annotations}
\label{fig:running-c}
 \end{figure} 

\paragraph{(1) C functions:} Our translation produces,
for each non-recursive rule definition in the RBR, a C
function without parameters that returns \code{void}.
Recursive rules produced by loops are translated into iterative code.  For this
  part of the translation, we compute the SCC from the CFG (see
  Fig.~\ref{tool}) and model the detected loops by means of
  \code{goto} instructions.
Fig.~\ref{fig:running-c} shows the obtained C functions from the RBR
program of Fig.~\ref{fig:running}. Note that $jump$ rules are
translated into an \emph{if-then-else} structure.

\paragraph{(2) Types of variables:} Solidity basic, signed and
unsigned data types are stored into untyped 256-bit words in the \EVM
bytecode, and the bytecode does not include information about the
actual types of the variables.  Moreover, most \EVM operations do not
distinguish among them except for few specific signed operations
(\code{SLT}, \code{SGT}, \code{SIGNEXTEND}, \code{SDIV} and
\code{SMOD}).  As 
verifiers behave differently w.r.t. overflow (see details in
\cite{DBLP:conf/cav/BeyerK11,Brockschmidt15,Kahsai15}), our
translation allows the user to choose (by means of a flag) if all
variables are declared with type \code{int} in the C program, or of
type \code{unsigned int} with casting to \code{int} for sign-specific
operations. The code in Fig.~\ref{fig:running-c} uses the default
\code{int} transformation.  Thus, although in \EVM integers have
overflow, the interpretation of them as unbounded integers or with
overflow will be determined by the available options in the C
verification tool (e.g., \verymax only handles unbounded integers).
Besides, instructions that contain \code{fresh}
variables or that are not handled (like \code{SLOAD}) are translated into a call
to function \code{\_\_VERIFIER\_nondet\_int} in order to model the lack of
information for them during verification. 
Observe that function \code{block734} includes some operations
over the different integer variables. 
Arrays or maps are not visible in the
\EVM (nor in the RBR). The only information that is trackable about arrays
corresponds to their sizes as it is stored in a stack variable that in the C
program is stored in an integer variable.

\paragraph{(3) Variable definitions:}
In order to enable reasoning on them (within their scopes) during verification,
\toolname translates them in the C program as follows:
(i) as we flattened the execution stack, we declare the stack variables as
global C variables to make them accessible to all C functions. 
These variables do not need to be  initialized as they take values in
the program code;
(ii) local variables are defined as global C variables
(L\ref{runc:local1}-L\ref{runc:local2}) because a function of the
contract might be translated into several C-functions, and all of them need
to access the local data. They are initialized at the beginning of the function
corresponding to the block in which they are firstly used;
(iii) state variables are also translated into global variables accessible by
all functions and, as their values when functions are verified are unknown, they
are initialized using  \code{\_\_VERIFIER\_nondet\_int}
(L\ref{runc:field1}-L\ref{runc:field2});
and (iv) function input parameters are also defined as global variables (for the
same reason as (ii)), whose initial values are not determined
(L\ref{runc:paramwho}).

\paragraph{(4)  SV-COMP annotations:}
The verification of Ethereum smart contracts is done in \toolname by
guaranteeing the unreachability of the \code{INVALID}
operations in the C-translated code. Following the \code{SV-COMP}
rules, we
translate \code{INVALID} operations into calls to
the \code{\_\_VERIFIER\_error} function so that its unreachability can be
proven by any verification tool compatible with the \code{SV-COMP} annotations.
An example of an \code{INVALID}  operation can be seen in L\ref{runc:error}.
%
Verification tools 
return that the program in Fig.~\ref{fig:running} cannot be verified
as the \code{INVALID} instruction could be
executed.
This is due to the fact that contract state values are unknown, that is:
\code{ttlSpply} is not guaranteed to be different from 0 at L\ref{run:div1}
and the size of the array \code{dvdnds} is not guaranteed to be greater than
the value of \code{lst} at L\ref{run:array}. Lines L\ref{run:req1} and
L\ref{run:req2} contain the \solidity instructions needed to guarantee that
L\ref{run:div1} and L\ref{run:array}, respectively,  will never
execute an \code{INVALID}
instruction. The \code{assert} at L\ref{run:assert} can be verified by using
the \code{require} at L\ref{run:req3}.
The inclusion of the \code{require} annotation also improves the
contract as, if it is violated, a \code{REVERT} rather than an
\code{INVALID} bytecode will be executed, not causing a loss of gas of
the transaction (while the gas needed to check it is negligible).





\section{Experimental Evaluation}\label{experiments}

All components of \toolname, except for the C verifiers, are
implemented in Python and are open-source. \toolname accepts smart
contracts written in versions of \solidity up to 0.4.25 and bytecode
for the Ethereum Virtual Machine v1.8.18. This section reports the
results of our experimental evaluation using \toolname with \cpa,
\seahorn and \verymax as verification back-ends. An artifact to try
our tool can be downloaded from
{\small\url{http://costa.fdi.ucm.es/papers/costa/safevm.ova}}.






In order to experimentally evaluate \toolname,
we pulled from \textsf{etherscan.io} all Ethereum contracts whose source
code was available on January 2018. This ended up in 10,796 files.
From those, we have searched for those files that contain \EVM code with
\code{INVALID} instructions, in total 7,323.
 The first phase
of \toolname that performs the decompilation into the RBR fails for
1,000 files (this 13.65\% is larger but quite aligned
with the failing rates of other
 tools e.g. \cite{Vandal,mythril}) and reaches a timeout of 60s for 22 files.
 Thus, our results are on the remaining 6,301 files, that 
 contain 24,294 contracts with 44,046 public functions that
 can reach an \code{INVALID} instruction and 177,549
 \code{INVALID}-free functions.
We have tested both  the translation to type \code{int} and
  \code{unsigned int} for defining C variables, as
 mentioned in
  Sec.~\ref{translator} for those 44,046 functions.
   We get the following results by using 60s of timeout
   (\textsf{Error} denotes an error output by the verifier):
\textcolor{white}{It is a fake line}




\begin{center}
\noindent
\begin{tabular}{|l@{\hspace{0.2cm}}||c@{\hspace{0.1cm}}|c@{\hspace{0.1cm}}||c@{\hspace{0.1cm}}|c@{\hspace{0.1cm}}||c@{\hspace{0.1cm}}|c@{\hspace{0.1cm}}|}
\hline
\multirow{2}{*}{\textbf{Results}} & \multicolumn{2}{c||}{\textbf{\cpa}} &
\multicolumn{2}{c||}{\textbf{\verymax}} &
\multicolumn{2}{c|}{\textbf{\seahorn}}\\\cline{2-7} 
& \code{int} & \code{uint} & \code{int} & \code{uint} & \code{int} &
\code{uint}\\
\hline
Verified & 19.48\%& 19.13\% & 20.32\%& 20.36\% & 21.71\%  &19.57\%   \\\hline

Non-Verified & 77.04\%& 79.82\% & 73.32\%& 73.44\% & 77.72\% &80.15\%\\\hline

Timeout & 3.21\%& 0.82\% & 6.29\%& 6.13\% & 0.57\% &0.28\% \\\hline

Error & 0.27\%& 0.23\% & 0.07\%& 0.07\% & 0\%  & 0\%\\\hline

\end{tabular}
\end{center}

\medskip

 The results for all verifiers are quite aligned, although \verymax verifies a
 slightly lower number of functions, and \seahorn
  verifies more functions and less
  reach a timeout.
 The interpretation made by the tools regarding the Integer
 semantics (bounded or unbounded) leads to the only relevant difference in the number of functions verified between both
 translations.

 We have manually inspected, out of the 7,323 files, those files whose
 addresses start with \texttt{0x00} and \texttt{0x01} in order to
 understand the cases that could not be verified. This is a sample of
 29 files (243 public functions) that are available at {\small
   \url{https://github.com/costa-group/EthIR/tree/master/examples/safevm}}.
 The manual inspection on the subset gives 54 false alarms (22.2\%),
 namely: 49 functions were verified by \cpa; 140 are correct alarms,
 most of them produced by asserts introduced by the programmers for
 safety to abort the execution (e.g. 83 come from \code{Safemath}); 54
 are false alarms (many related to \code{enum} accesses and other
 imprecisions in the decompilation phase).
 More in detail, we have identified four types of situations: (1)
 false alarms due to \emph{inaccuracy of our tool}: some \code{assert}
 statements contain non-integer types (e.g., strings, \code{enum},
 etc.) 
 which cannot be verified as we need a more accurate decompilation
 (see Sec.~\ref{related}); (2) correct alarms that require
 \emph{conditional verification}: some \code{assert} statements can
 only be verified for concrete contexts, e.g., we found \code{assert}s
 to prevent from under/overflow integer arithmetic operations in a
 widely used library \code{SafeMath} that can only be verified for
 given inputs. In the future we plan to integrate conditional
 verification \cite{Brockschmidt15} to infer the preconditions for the
 \code{assert}s to hold;
 (3) Correct alarms detecting \emph{potential vulnerabilities}: we
 have detected several \code{INVALID} operations that could represent
 a vulnerability in the code (e.g., functions that access an array
 element without checking the boundary) and we have protected them
 adding \code{require} statements that enable subsequent verification;
 and (4) four functions whose verification results depend on the
 different semantics used for Integers.

 As final observations, we notice that \code{assert} is overused
 (contradicting the best practices recommendations of \solidity) and
 that some contracts can be improved by using \code{require} to avoid
 the loss of gas when the \code{assert} statement does not hold.
 Finally, we argue that although there is much room for improving the
 accuracy, the results of our experimental evaluation are very
 encouraging: we have verified safety w.r.t.\ \code{INVALID} bytecodes
 for around 20\% of the functions that might reach \code{INVALID}
 fully automatically by using state-of-the-art verifiers.

\section{Conclusions}\label{related}

 Verification of Ethereum smart contracts for
potential safety and security vulnerabilities is becoming a popular
research topic with numerous tools being developed, among them, we
have tools based on symbolic
execution~\cite{Luu-al:CCS16,GrossmanAGMRSZ18,Nikolic-al:Maian,KruppR18,Kalra-al:NDSS18,TsankovDDGBV18},
tools based on SMT
solving~\cite{Marescotti-al:ISoLA18,Kolluri-al:laws}, and other tools based
on certified
programming~\cite{Bhargavan-al:PLAS16,Grishchenko-al:POST18,Amani-al:CPP18}.
There are some tools also that aim at detecting, analyzing and
verifying non-functional properties of smart contracts, e.g., those
focused on reasoning about the gas consumption
\cite{AlbertGRS19,ChenLLZ17,madmax,Marescotti-al:ISoLA18}.

To the best of our knowledge, \toolname is the first tool that uses
existing verification engines developed for C programs to verify
low-level \EVM code. This opens the door to the applicability of
advanced techniques developed for the verification of C programs to
the new languages used to code smart contracts.
Although our tool is still in a prototypical stage, it provides a
proof-of-concept of the transformational approach, and we argue that
it constitutes a promising basis to build verification tools for \EVM
smart contracts. Some of the aspects that we aim at improving in
future work is the handling of the data stored in the memory, as it is
abstracted away by the \ethir component that \toolname is using as
soon as there are storage operations on memory. Developing a memory
analysis for \EVM smart contracts can be crucial for the accuracy of
verification. We also aim at handling bit-wise operations in the
future that are extensively used in the \EVM bytecode. Advanced
reasoning for arrays and maps (the only data structures available in
Ethereum smart contracts) can be also added to the framework to gain
further accuracy. This requires also further work on the decompilation
side. Along the same line, learning information on the types of
variables during decompilation will have an impact in the accuracy of
the verification process.

\vspace{-0.3cm}
\section*{Acknowledgments}
This work was funded partially by the Spanish MINECO project
TIN2015-69175-C4-2-R and MINECO/FEDER, UE project
TIN2015-69175-C4-3-R, by Spanish MICINN/FEDER, UE projects
RTI2018-094403-B-C31 and RTI2018-094403-B-C33, by the CM projects
S2018/\-TCS-4314 and S2018/TCS-4339, co-funded by EIE Funds of the
European Union, and by the UCM CT27/16-CT28/16 grant.

\bibliographystyle{plain}
\bibliography{biblio}

\end{document}